\documentclass{article}
\usepackage{spconf,amsmath,amssymb,graphicx,xcolor,url,hyperref}
\usepackage{enumitem}
\usepackage{multirow}
\usepackage{hhline}
\setlist{nosep, leftmargin=14pt}
\usepackage[percent]{overpic}
\usepackage{mwe} 
\usepackage{color}
\usepackage{mdframed}
\usepackage{blindtext}
\usepackage[export]{adjustbox}

\definecolor{blau}{RGB}{122 172 172} 
\definecolor{rot}{RGB}{ 172,122,122 } 
\title{Prediction of low-keV monochromatic images from polyenergetic CT scans for improved automatic detection of pulmonary embolism}
\name{\begin{tabular}{c}Constantin Seibold$^{*1}$, Matthias A. Fink$^{*2}$,Charlotte Goos$^1$, Hans-Ulrich Kauczor$^2$ \\
$\textit{Heinz-Peter Schlemmer}^3$, Rainer Stiefelhagen$^1$, Jens Kleesiek$^{3,4}$\end{tabular}
\thanks{$^*$ denotes equal contribution.}
}

\address{$^1$ Institute of Anthropomatics \& Robotics, Karlsruhe Institute of Technology, Germany\\ $^2$
Department of Diagnostic and Interventional Radiology, University Hospital Heidelberg, Germany\\$^3$
German Cancer Research Center, Heidelberg, Germany\\ $^4$Institute for AI in Medicine (IKIM), University Hospital Essen, Germany}

\begin{document}

\maketitle

\begin{abstract}
Detector-based spectral computed tomography is a recent dual-energy CT (DECT) technology that offers the possibility of obtaining spectral information. From this spectral data, different types of images can be derived, amongst others virtual monoenergetic (\textit{monoE}) images. \textit{MonoE} images potentially exhibit decreased artifacts, improve contrast, and overall contain lower noise values, making them ideal candidates for better delineation and thus improved diagnostic accuracy of vascular abnormalities.

In this paper, we are training convolutional neural networks~(CNN) that can emulate the generation of \textit{monoE} images from conventional single energy CT acquisitions. For this task, we investigate several commonly used image-translation methods. We demonstrate that these methods while creating visually similar outputs, lead to a poorer performance when used for automatic classification of pulmonary embolism (PE). We expand on these methods through the use of a multi-task optimization approach, under which the networks achieve improved classification as well as generation results, as reflected by PSNR and SSIM scores. Further, evaluating our proposed framework on a subset of the RSNA-PE challenge data set shows that we are able to improve the Area under the Receiver Operating Characteristic curve (AuROC) in comparison to a naïve classification approach from 0.8142 to 0.8420.

\end{abstract}
\section{Introduction}
\label{sec:intro}
In spectral computed tomography (DECT), projection data simultaneously obtained from both detector layers is utilized to generate spectral images such as virtual monoenergetic (\textit{monoE}) scans. Next to the conventional (polyenergetic) images, multiple spectrally distinct attenuation maps can be obtained from a single scan and used to derive different types of images.
The clinical uses of the DECTs can be summarized with enhanced visualization of intravascular contrast, reduction of artifacts such as calcium blooming, material decomposition, and radiation dose reduction~\cite{spectral}. Therefore, in comparison to conventional CT, DECT compares favorably for the diagnosis of various diseases such as myocardial perfusion~\cite{myo_study} or pulmonary embolisms~\cite{pe_study}.

We argue that similar to the expert radiologist, convolutional neural networks (CNN) may benefit when trained on DECT data. However, as most currently existing CT data sets were acquired with conventional CT scanners they do not comprise \textit{monoE} images. To bridge this gap, we investigate the use of existing image-translation models such as Pix2Pix~\cite{pix2pix}, which might be able to use the underlying distributions in polyenergetic images to predict spectral images akin to what was done to translate CT images to MRI scans~\cite{wolterink2017deep}. In turn, these generated synthetic \textit{monoE} images might be used as input for CNNs potentially facilitating their detection of pathologies.
While existing image-translation methods are able to generate visually appealing results they do not enforce features that enable the correct identification of certain classes. For this reason, we introduce a joint optimization between the generation of the monoenergetic domain and the simultaneous identification of pathologies. This leads to a network that learns to combine features necessary for a downstream classification task as well as for synthetic image generation. In other words, the proposed framework learns a suitable mapping on the basis of monoenergetic images.
\def\arraystretch{0.75}\tabcolsep=5pt

\begin{figure*}[t]
    \centering
    \begin{tabular}{c|c}
        \frame{\begin{overpic}[height=0.09\textheight]{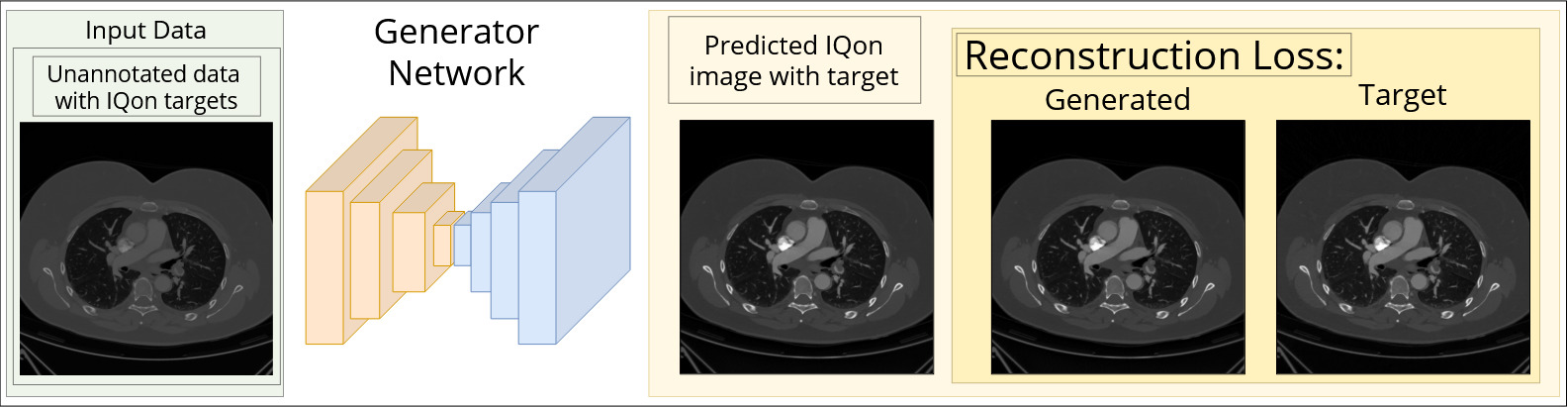}
         \put (1,23) {\small a)} 
            \end{overpic}}
        &  \multirow{2}{*}[52pt]{\frame{\begin{overpic}[height=0.20\textheight,width=0.52\textwidth]{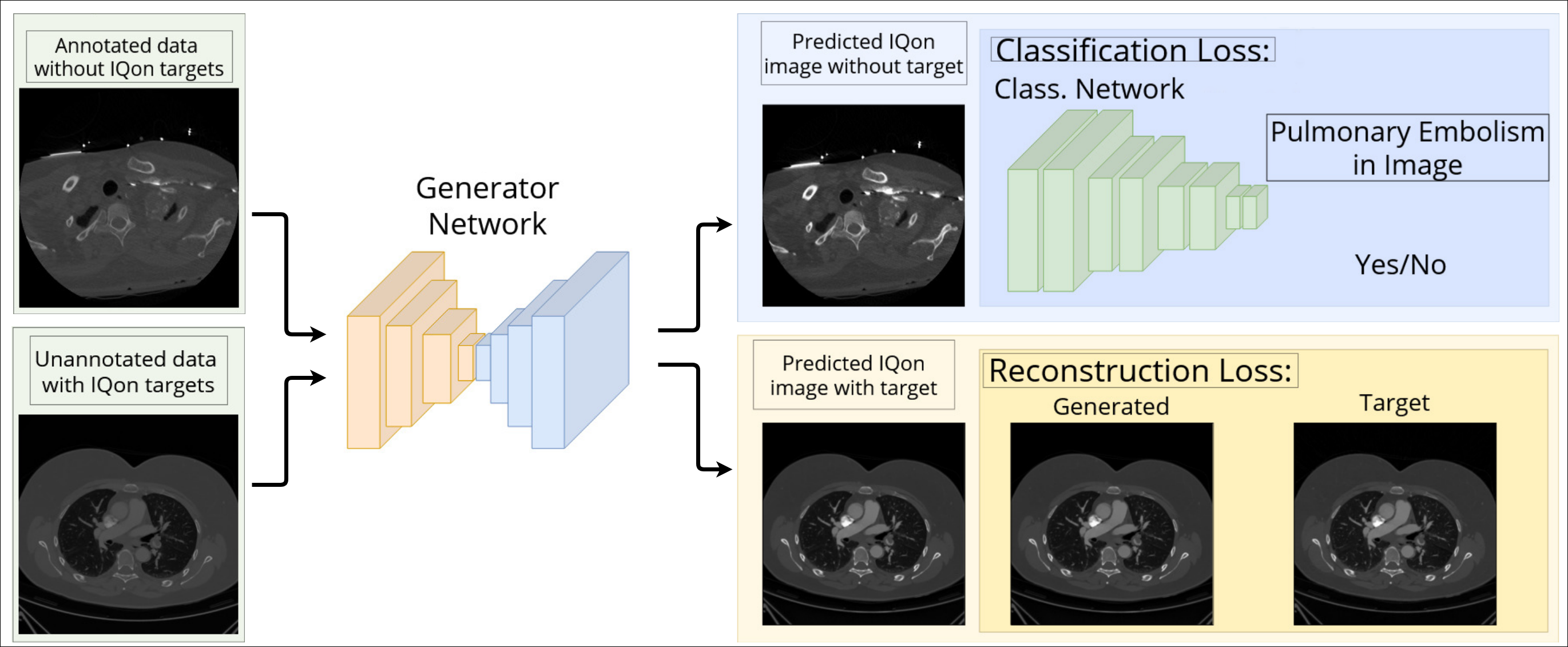}\put (1,47) {\small c)} \end{overpic}}} \\\cline{1-1}\\
        \frame{\begin{overpic}[height =0.09\textheight]{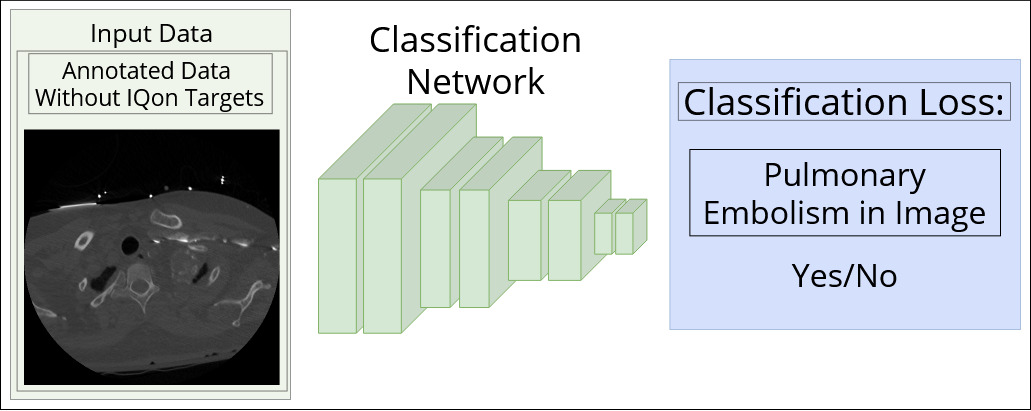}
         \put (1,35) {\small b)} 
            \end{overpic}}
        
    \end{tabular}
    
    \caption{Overview of the different approaches for the combination of domain adaptation and classification. On the left, a) trains a generator on the paired dataset, which would be followed by the training of a classifier on annotated data in b). c) displays our approach of the joint optimization of generator and classifier, where the generator learns the mapping for unlabeled data, while adjusting its features in a way that it does not hinder the classification network when given annotated data.}
    \label{fig:approach}
\end{figure*}
Our contributions can be summarized as (1) an extensive study comparing various image-translation methods for the prediction of \textit{monoE} images from conventional polyenergetic scans, (2) evaluation of the classification accuracy of predicted synthetic \textit{monoE} images for the detection of PE, and (3) proposal of a training regime, enabling to generate data that is not only visually similar but also incorporates features necessary for the automatic identification of pathologies.

\section{Methods and Materials}
\label{sec:methods}

Suppose, we are given two distinct data sets $D_1$ and $D_2$. $D_1$ consists of unannotated images with poly- and monoenergetic depictions. $D_2$ describes a set of images with slice-level disease annotations without corresponding monoenergetic representation.  We now aim to design a unified model that jointly optimizes disease identification and domain adaption most fitting for the task. We have formulated these two tasks into the same framework so that 1) it trains these tasks end-to-end and 2) the two tasks can be mutually beneficial. The proposed architecture is displayed in Fig.~\ref{fig:approach}~c).

\noindent\textbf{Methodology:}
The proposed framework jointly optimizes two tasks in an end-to-end manner. As one task, we consider the problem of translating between the domain of polyenergetic $x \in X$ and monoE images $y\in Y$ as a paired image-translation problem.  Here, a generator aims to learn a mapping $G:x\rightarrow y$, which minimizes the difference between the two paired images.
This objective can be expressed as 
\begin{equation}
    \mathcal{L}_{L1} = \mathbb{E}_{x,y} [||G(x)-y||_1].
\end{equation}
We utilize the mean absolute error as it has been found to lead to less blurry images~\cite{pix2pix}.

Consecutively, the output of the generator is fed into a classification network $C$, which attempts to predict the occurrence of a disease label $z$, $C: G(x) \rightarrow z$ of the annotated data set. We utilize ResNet50~\cite{he2016deep}, however our framework can be easily extended to employ any other existing CNN model. We utilize a sigmoid activation $\sigma$ for making output predictions. 
\begin{equation} \label{eq1}
\begin{split}
    \mathcal{L}_{cls} = &\mathbb{E}_{x,z} [- z \log  \sigma(C(G(x)))\\& - (1-z) \log (1-\sigma(C(G(x)))]
\end{split}
\end{equation}
To optimize both objectives during the training process, we construct our data set as a combination of the two data sets (see below) and sample the batch in a way such that on average it consists of 50\% of either. Therefore, target disease labels are only given for half of the batch and monoenergetic target images for the other half. To accommodate this circumstance into the optimization function we introduce a marker variable $m$, which switches between $[0,1]$ depending on whether we are presented a target image $y$ or a target label $z$. In this manner, the final loss can be formulated as
\begin{equation}
    \mathcal{L} = m * L_{cls} + (1-m) * L_{L1}.
\end{equation}
For backpropagation of the gradients one network is frozen, while the other is updated similar to an adversarial training. 

\noindent\textbf{Implementation Details:}
We train our networks jointly in an end-to-end manner by sequentially passing data through the generator and classification network. Our generator network utilizes a fully convolutional 9-block ResNet encoder-decoder network, however, similar to our classifier, the model  can easily be replaced by more advanced architectures. We use Adam for optimization with a learning rate of 0.0002, $\beta_1=0.9$ and $\beta_2=0.99$ with a weight decay of 0.00001. After training for 5 epochs on the joint data set, we decay our learning rate to 0 over the following 5 epochs. We use an image-size of $512\times512$ with a batch size of 5 for all our experiments. 
\def\arraystretch{1}\tabcolsep=1.5pt
\setlength{\abovecaptionskip}{1pt}
\setlength{\belowcaptionskip}{-30pt}

\begin{figure*}[t]
    \centering
    \begin{tabular}{cccccccc}
         \hline Input & L1 & SPL & Pix2Pix & CRN & Pix2PixHD & Ours &Target  \\\hline\vspace{-0.2cm}\\
         \frame{\begin{overpic}
            [width=0.11\textwidth]{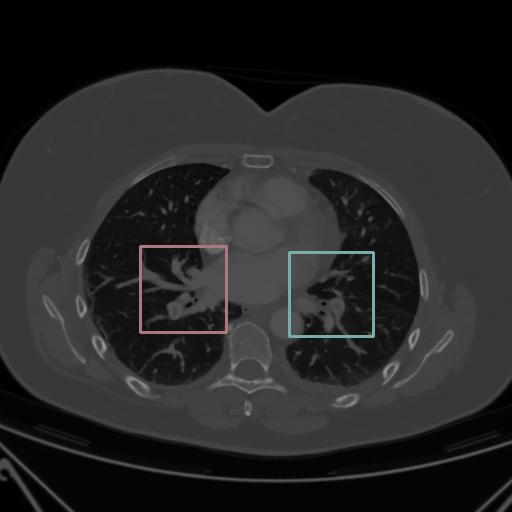} \put (25,93) {\tiny \textcolor{white}{$SSIM:0.934$}} 
            \put (23,84) {\tiny \textcolor{white}{$PSNR:28.16$}}
         \end{overpic}}
         
         & \frame{\begin{overpic}[width=0.11\textwidth]{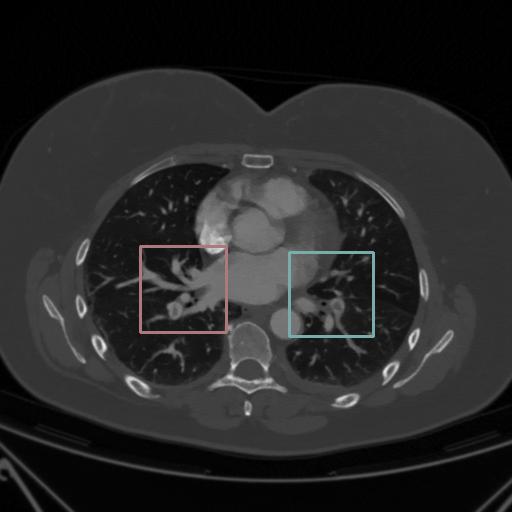}\put (25,93) {\tiny \textcolor{white}{$SSIM:0.989$}} 
            \put (23,84) {\tiny \textcolor{white}{$PSNR:43.40$}}
            \end{overpic}} 
         & \frame{\begin{overpic}[width=0.11\textwidth]{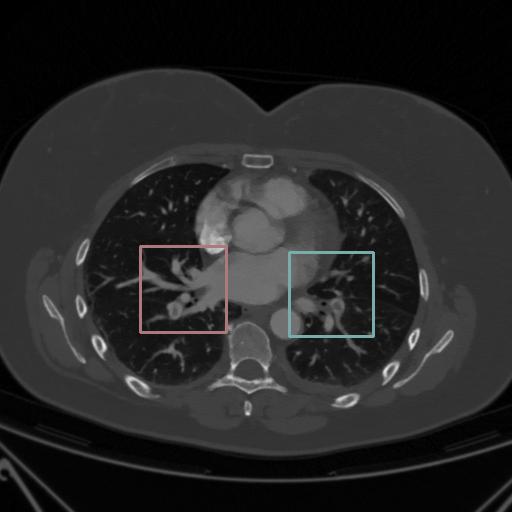}\put (25,93) {\tiny \textcolor{white}{$SSIM:0.987$}} 
            \put (23,84) {\tiny \textcolor{white}{$PSNR:41.24$}}
            \end{overpic}} 
         & \frame{\begin{overpic}[width=0.11\textwidth]{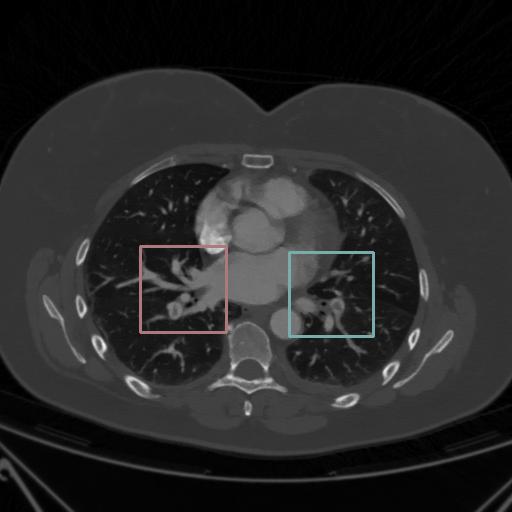} \put (25,93) {\tiny \textcolor{white}{$SSIM:0.986$}} 
            \put (23,84) {\tiny \textcolor{white}{$PSNR:42.45$}}
         \end{overpic}} 
         & \frame{\begin{overpic}[width=0.11\textwidth]{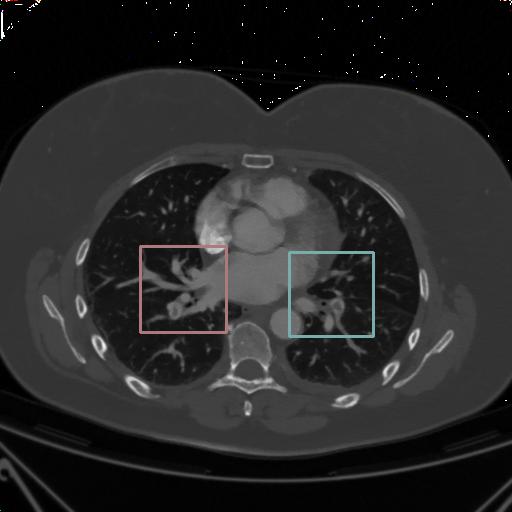}\put (25,93) {\tiny \textcolor{white}{$SSIM:0.986$}} 
            \put (23,84) {\tiny \textcolor{white}{$PSNR:39.73$}}
            \end{overpic}} 
         & \frame{\begin{overpic}[width=0.11\textwidth]{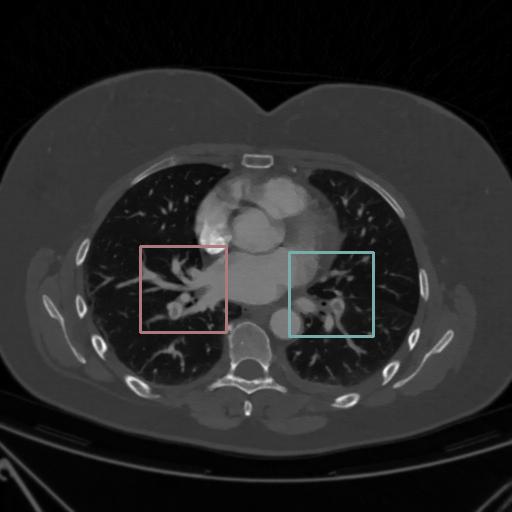}\put (25,93) {\tiny \textcolor{white}{$SSIM:0.972$}} 
            \put (23,84) {\tiny \textcolor{white}{$PSNR:37.15$}}
            \end{overpic}} 
            & \frame{\begin{overpic}[width=0.11\textwidth]{figures/images/comparison/69278670_generated_pix2pixl1.jpg}\put (25,93) {\tiny \textcolor{white}{$SSIM:0.988$}} 
            \put (23,84) {\tiny \textcolor{white}{$PSNR:43.46$}}
            \end{overpic}} 
         & \frame{\begin{overpic}[width=0.11\textwidth]{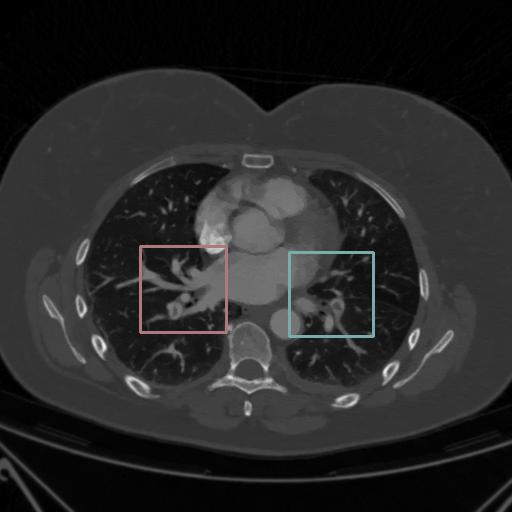}\put (25,93) {\tiny \textcolor{white}{$SSIM:1.000$}} 
            \put (23,84) {\tiny \textcolor{white}{$PSNR: - $}}
            \end{overpic}} 
         \\ \vspace{-0.495cm} \\\begin{tabular}{cc}

              \includegraphics[width=0.05\textwidth,cframe=rot 0.75pt]{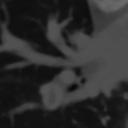}\hspace{-1.25pt}
              & \includegraphics[width=0.05\textwidth,cframe=blau 0.75pt]{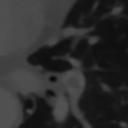}
            \end{tabular}
            & \begin{tabular}{cc}
              \includegraphics[width=0.05\textwidth,cframe=rot 0.75pt]{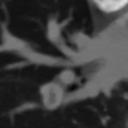}\hspace{-1.25pt}
              & \includegraphics[width=0.05\textwidth,cframe=blau 0.75pt]{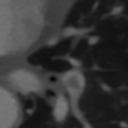}
            \end{tabular}
            & \begin{tabular}{cc}
              \includegraphics[width=0.05\textwidth,cframe=rot 0.75pt]{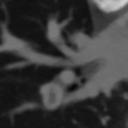}\hspace{-1.25pt}
              & \includegraphics[width=0.05\textwidth,cframe=blau 0.75pt]{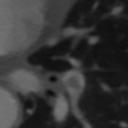}
            \end{tabular}
            & \begin{tabular}{cc}
              \includegraphics[width=0.05\textwidth,cframe=rot 0.75pt]{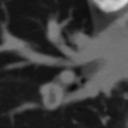}\hspace{-1.25pt}
              & \includegraphics[width=0.05\textwidth,cframe=blau 0.75pt]{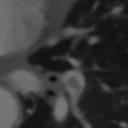}
            \end{tabular}
            & \begin{tabular}{cc}
              \includegraphics[width=0.05\textwidth,cframe=rot 0.75pt]{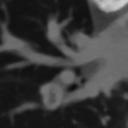}\hspace{-1.25pt}
              & \includegraphics[width=0.05\textwidth,cframe=blau 0.75pt]{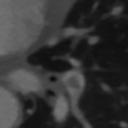}
            \end{tabular}
            & \begin{tabular}{cc}
              \includegraphics[width=0.05\textwidth,cframe=rot 0.75pt]{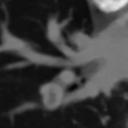}\hspace{-1.25pt}
              & \includegraphics[width=0.05\textwidth,cframe=blau 0.75pt]{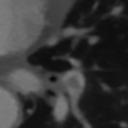}
            \end{tabular}
            & \begin{tabular}{cc}
              \includegraphics[width=0.05\textwidth,cframe=rot 0.75pt ]{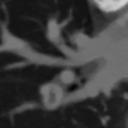}\hspace{-1.25pt}
              & \includegraphics[width=0.05\textwidth,cframe=blau 0.75pt]{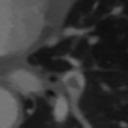}  
            \end{tabular}
        & \begin{tabular}{cc}
              \includegraphics[width=0.05\textwidth,cframe=rot 0.75pt ]{figures/images/comparison/69278670_target_2_snippet.jpg}\hspace{-1.25pt}
              & \includegraphics[width=0.05\textwidth,cframe=blau 0.75pt]{figures/images/comparison/69278670_target_1_snippet.jpg}  
            \end{tabular}
            
        \\
        \vspace{-0.33cm}
        \\
        \frame{\begin{overpic}[width=0.11\textwidth]{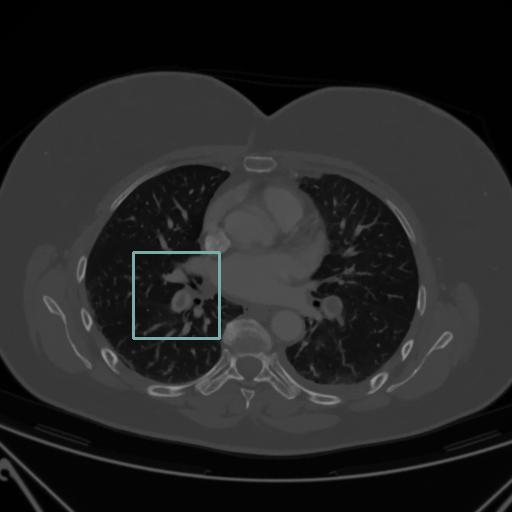}
        \put (25,93) {\tiny \textcolor{white}{$SSIM:0.932$}} 
            \put (23,84) {\tiny \textcolor{white}{$PSNR:27.89$}}
            \end{overpic}}\llap{\includegraphics[height=0.65cm,cframe=blau 0.75pt]{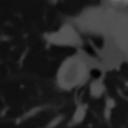}}
         & \frame{\begin{overpic}[width=0.11\textwidth]{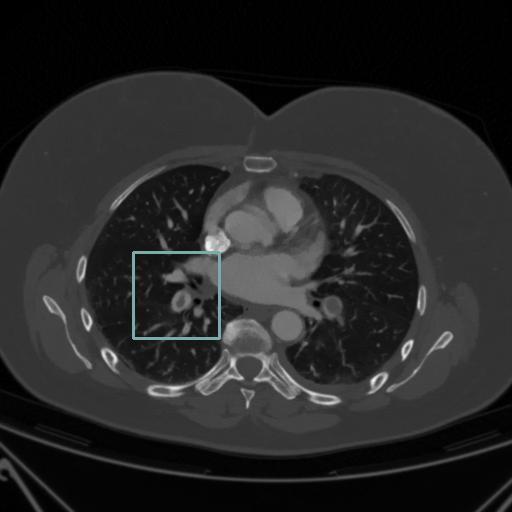}\put (25,93) {\tiny \textcolor{white}{$SSIM:0.987$}} 
            \put (23,84) {\tiny \textcolor{white}{$PSNR:42.98$}}
            \end{overpic}}\llap{\includegraphics[height=0.65cm,cframe=blau 0.75pt]{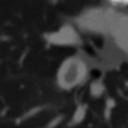}}
         &\frame{\begin{overpic}[width=0.11\textwidth]{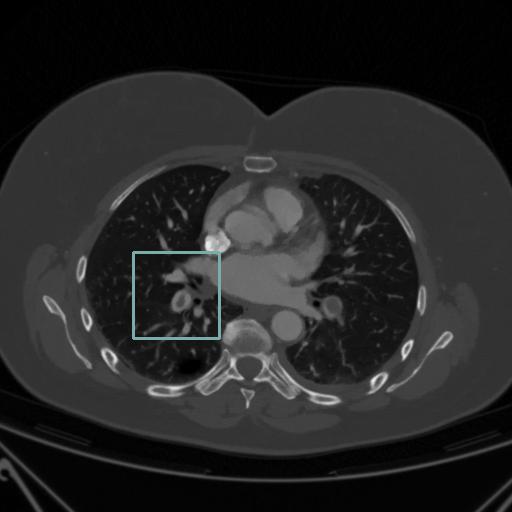}\put (25,93) {\tiny \textcolor{white}{$SSIM:0.984$}} 
            \put (23,84) {\tiny \textcolor{white}{$PSNR:39.96$}}\end{overpic}}\llap{{\includegraphics[height=0.65cm,cframe=blau 0.75pt]{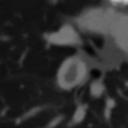}}}
         &\frame{\begin{overpic}[width=0.11\textwidth]{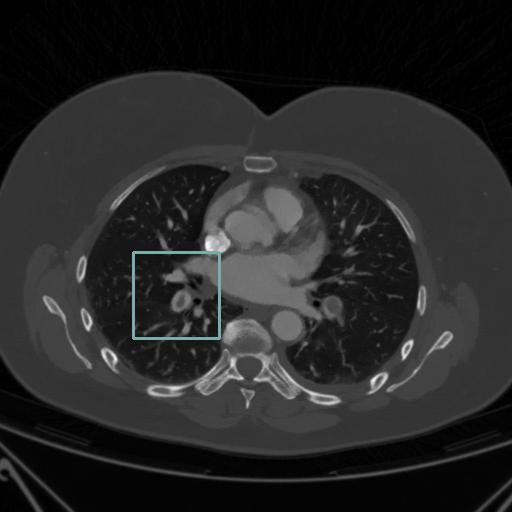}\put (25,93) {\tiny \textcolor{white}{$SSIM:0.984$}} 
            \put (23,84) {\tiny \textcolor{white}{$PSNR:42.065$}}
            \end{overpic}}\llap{\includegraphics[height=0.65cm,cframe=blau 0.75pt]{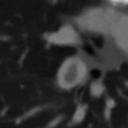}}
         & \frame{\begin{overpic}[width=0.11\textwidth]{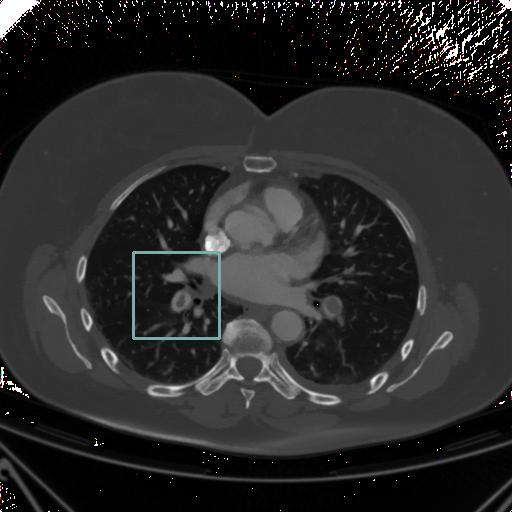}\put (25,93) {\tiny \textcolor{white}{$SSIM:0.985$}} 
            \put (23,84) {\tiny \textcolor{white}{$PSNR:40.325$}}
            \end{overpic}}\llap{\includegraphics[height=0.65cm,cframe=blau 0.75pt]{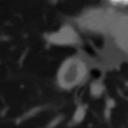}} 
         & \frame{\begin{overpic}[width=0.11\textwidth]{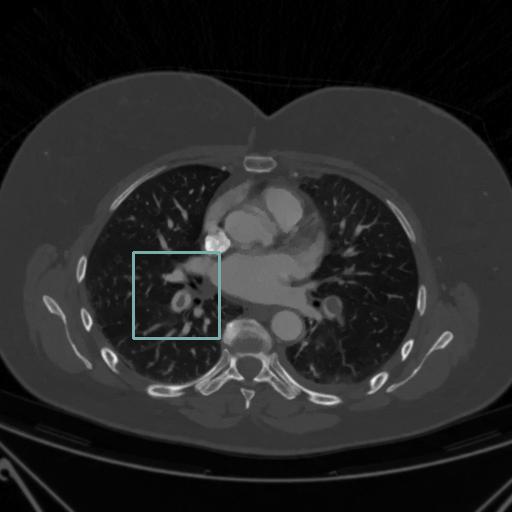}\put (25,93) {\tiny \textcolor{white}{$SSIM:0.972$}} 
            \put (23,84) {\tiny \textcolor{white}{$PSNR:37.70$}}
            \end{overpic}}\llap{\includegraphics[height=0.65cm,cframe=blau 0.75pt]{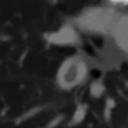}} 
         & \frame{\begin{overpic}[width=0.11\textwidth]{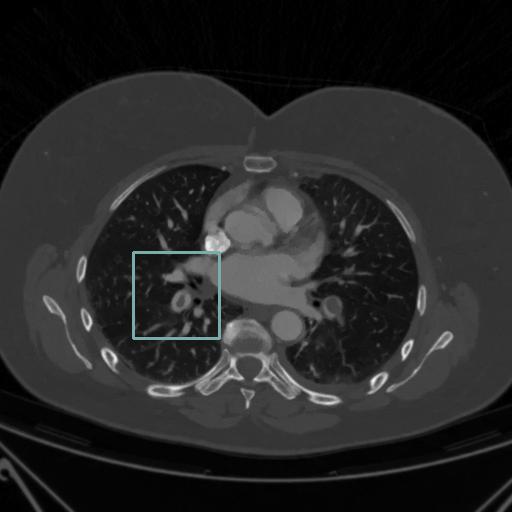}\put (25,93) {\tiny \textcolor{white}{$SSIM:0.986$}} 
            \put (23,84) {\tiny \textcolor{white}{$PSNR:42.48$}}
            \end{overpic}}\llap{\includegraphics[height=0.65cm,cframe=blau 0.75pt]{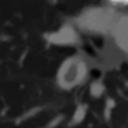}}
        & \frame{\begin{overpic}[width=0.11\textwidth]{figures/images/comparison/69278549_target.jpg}\put (25,93) {\tiny \textcolor{white}{$SSIM:1.000$}} 
            \put (23,84) {\tiny \textcolor{white}{$PSNR:-$}}
            \end{overpic}}\llap{\includegraphics[height=0.65cm,cframe=blau 0.75pt]{figures/images/comparison/69278549_target_1_snippet.jpg}}
         \\
        \vspace{-0.33cm}
        \\
        \frame{\begin{overpic}[width=0.11\textwidth]{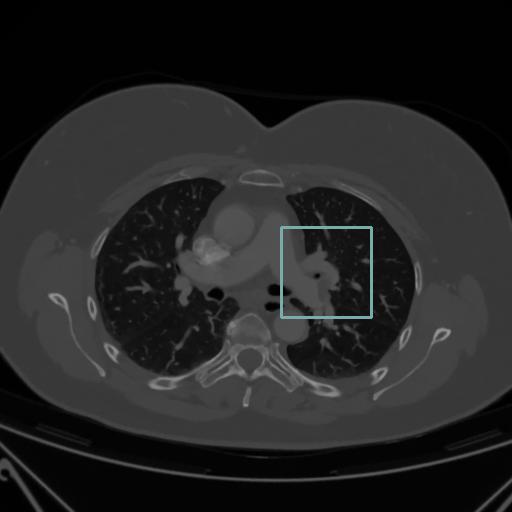}
        \put (25,93) {\tiny \textcolor{white}{$SSIM:0.927$}} 
            \put (23,84) {\tiny \textcolor{white}{$PSNR:28.34$}}
            \end{overpic}}\llap{\includegraphics[height=0.65cm,cframe=blau 0.75pt]{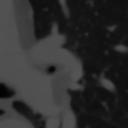}}
         &\frame{\begin{overpic}[width=0.11\textwidth]{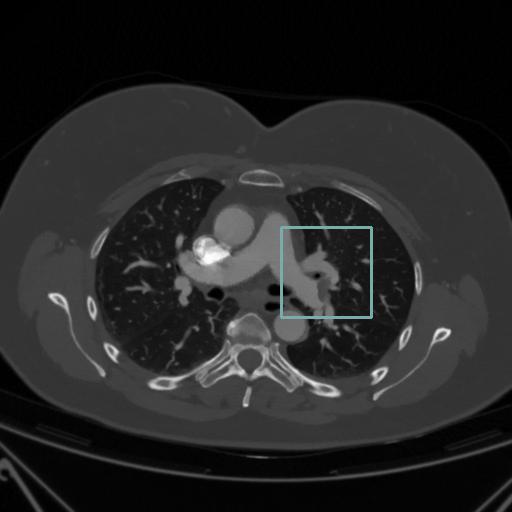}
         \put (25,93) {\tiny \textcolor{white}{$SSIM:0.989$}} 
            \put (23,84) {\tiny \textcolor{white}{$PSNR:43.62$}}
            \end{overpic}}\llap{\includegraphics[height=0.65cm,cframe=blau 0.75pt]{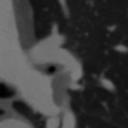}}
         &\frame{\begin{overpic}[width=0.11\textwidth]{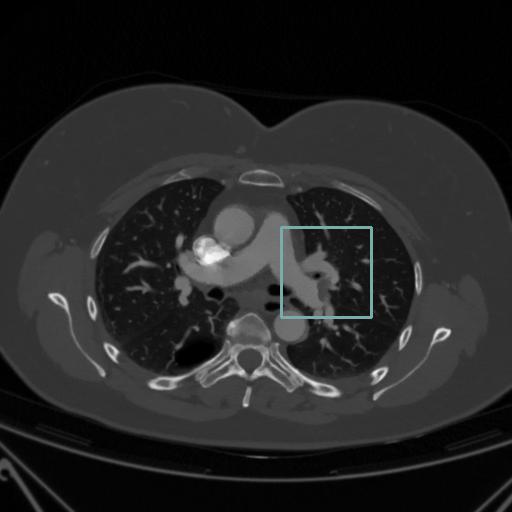}
         \put (25,93) {\tiny \textcolor{white}{$SSIM:0.984$}} 
            \put (23,84) {\tiny \textcolor{white}{$PSNR:39.11$}}
            \end{overpic}}\llap{\includegraphics[height=0.65cm,cframe=blau 0.75pt]{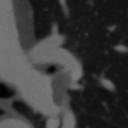}}
         & \frame{\begin{overpic}[width=0.11\textwidth]{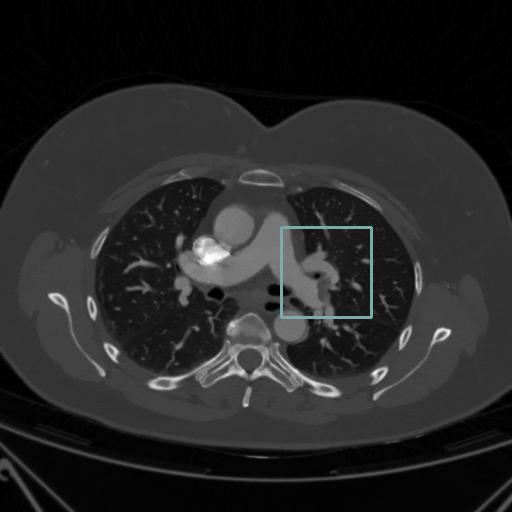}
         \put (25,93) {\tiny \textcolor{white}{$SSIM:0.986$}} 
            \put (23,84) {\tiny \textcolor{white}{$PSNR:42.96$}}
            \end{overpic}}\llap{\includegraphics[height=0.65cm,cframe=blau 0.75pt]{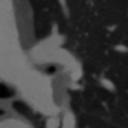}}
         & \frame{\begin{overpic}[width=0.11\textwidth]{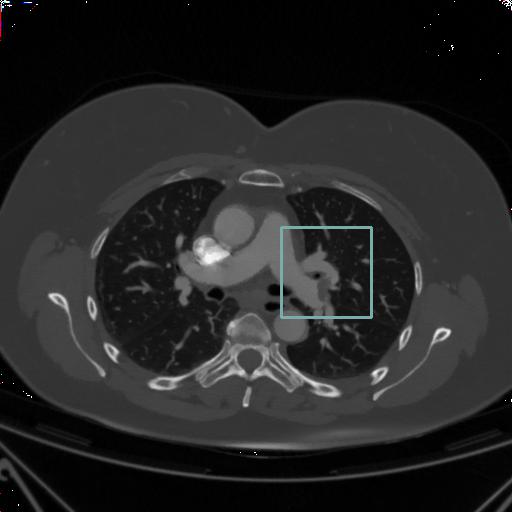}
         \put (25,93) {\tiny \textcolor{white}{$SSIM:0.986$}} 
            \put (23,84) {\tiny \textcolor{white}{$PSNR:40.11$}}
            \end{overpic}}\llap{\includegraphics[height=0.65cm,cframe=blau 0.75pt]{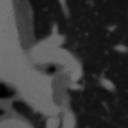}} 
         & \frame{\begin{overpic}[width=0.11\textwidth]{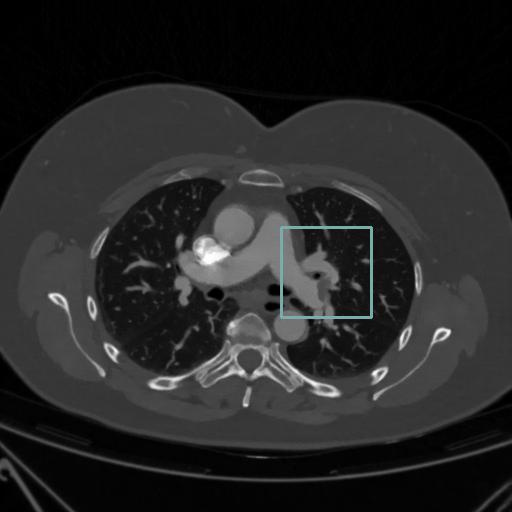}
         \put (25,93) {\tiny \textcolor{white}{$SSIM:0.978$}} 
            \put (23,84) {\tiny \textcolor{white}{$PSNR:39.55$}}
            \end{overpic}}\llap{\includegraphics[height=0.65cm,cframe=blau 0.75pt]{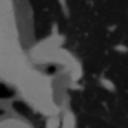}} 
         & \frame{\begin{overpic}[width=0.11\textwidth]{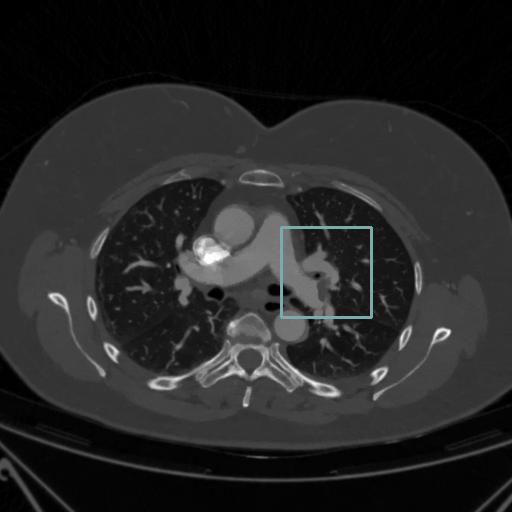}
         \put (25,93) {\tiny \textcolor{white}{$SSIM:0.982$}} 
            \put (23,84) {\tiny \textcolor{white}{$PSNR: 41.87$}}
            \end{overpic}}\llap{\includegraphics[height=0.65cm,cframe=blau 0.75pt]{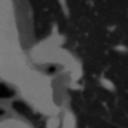}}
          & \frame{\begin{overpic}[width=0.11\textwidth]{figures/images/comparison/69278219_target.jpg}
         \put (25,93) {\tiny \textcolor{white}{$SSIM:1.000$}} 
            \put (23,84) {\tiny \textcolor{white}{$PSNR: -$}}
            \end{overpic}}\llap{\includegraphics[height=0.65cm,cframe=blau 0.75pt]{figures/images/comparison/69278219_target_1_snippet.jpg}}
    \end{tabular}
    \caption{Qualitative comparison of different image translation methods on our internal DE-CTPA dataset. Individual SSIM and PSNR values are shown on the images. Areas around \textit{Pulmonary Embolisms} are displayed seperately.}
    \label{fig:comparison}
\end{figure*}

\noindent\textbf{Experimental Setup:} \label{data_set}
We utilize two data sets for our experiments. 
The private dual-energy computed tomography pulmonary angiography (DE-CTPA) data set $D_1$ was gathered during routine clinical workup of 27 consecutive patients with suspected pulmonary embolism. 
The CT scans were performed on a dual-layer detector (IQon Spectral CT, Philips Healthcare). Standard arterial series and the corresponding monoenergetic images at low-energy levels (40 keV) were reconstructed. The data set contains 7892 image pairs.

The second data set $D_2$ is a subset of the \textit{RSNA STR Pulmonary Embolism Detection}~\cite{rsna}. Out of the annotated 7279 subjects, we sample 10\% of the training data patient-wise. The sampled data set consists of a total of 161253 annotated slices with roughly the same label distribution as present in the open training set. We further split the data patient-wise 50\%/25\%/25\% into train-, val- and test-sets, respectively. 

For our experiments on our DE-CTPA, we perform 5-fold cross validation and average our reconstruction results in terms of Peak-Signal-to-Noise-Ratio (PSNR) and Structural Similarity Index Measure (SSIM). 
For the identification of PE, we perform binary classification on slice level for each presented image domain and report the area under the receiver-operating-characteristic (AuROC) on the test split of the model which performed best on the validation set. We validated our model after each epoch. 

We compare against various image-translation models such as an L1-loss-based generator as a baseline, Pix2Pix\cite{pix2pix}, Pix2PixHD\cite{pix2pixhd}, CRN\cite{crn}, SPL\cite{spl}. We further added L1-Losses to feature-loss based methods (CRN, Pix2PixHD), denoted by *. All methods apart from CRN and Pix2PixHD, which use their originally proposed architecture, are trained using the same 9-block ResNet architecture. \textit{Orig.} denotes the direct usage of conventional CT imagery for either the computation of PSNR/SSIM or as input into a classification network.
To evaluate classification performance for different image translation methods, we train all methods on the same split in the cross validation setting of our internal data set.

 \noindent\textbf{Compliance with ethical standards:}
 The first data set was gathered as part of a retrospective single-centre HIPAA-compliant study, which was approved by the local institutional review board (No. S-236/2020) with a waiver for written informed consent.
 As the second data set is part of a public competition ethical approval was not required as confirmed by the license attached with the open access data.
 
 \noindent\textbf{Conflicts of Interest:}
 No funding was received for conducting this study. The authors have no relevant financial or non-financial interests to disclose.

\section{Results}
\label{sec:Results}
    \setlength{\abovecaptionskip}{1pt}
    \setlength{\belowcaptionskip}{-20pt}
\begin{table}[t]
    \centering
    \begin{tabular}{|c||c|c|}
        \hline   Method &  SSIM & PSNR \\\hline\hline
        Orig. & $ 0.945 \pm 0.007 $& $ 30.189 \pm  0.690 $  \\\hline\hline
        L1 & $\textbf{ 0.984} \pm \textbf{0.002	}$ & $\textbf{ 	42.365} \pm \textbf{0.642 }$  \\\hline
        SPL\cite{spl} & \textit{ 0.983} $\pm$ \textit{0.002 } & $ 40.888 \pm  0.216 $ \\\hline\hline
        Pix2Pix~\cite{pix2pix} &  $ 0.978 \pm  0.003 $ &  $ 40.897 \pm  0.697 $ \\\hline
        Pix2PixHD~\cite{pix2pixhd} &  $ 0.971 \pm  0.004 $ &  $ 38.739 \pm  0.624 $ \\\hline
        CRN~\cite{crn} &  $ 0.371 \pm  0.551 $&  $ 19.482 \pm  16.033 $\\\hline\hline
        Pix2PixHD* &  $ 0.971 \pm  0.004 $ &  $ 38.415 \pm  1.278 $ \\\hline
        CRN* &  $ 0.976 \pm  0.0045 $&  $ 37.582 \pm  1.574 $\\\hline\hline
        Ours & $\textbf{0.984}\pm \textbf{0.002}$&$ \textit{41.706} \pm \textit{0.547}$\\\hline

    \end{tabular}
    \caption{Reconstruction results of various Image-Translation methods. Best and second-best result in $\textbf{bold}$ and  $\textit{cursive}$.}
    \label{tab:psnr_ssim}
\end{table}
\noindent\textbf{Quantative Results of Translation Properties:}
\noindent The quantitative results on the reconstruction ability are displayed in Table~\ref{tab:psnr_ssim}. Models optimized on image-based comparison outperform feature-loss and adversarial methods for the evaluated task. Our method achieves similar performance to the L1-based generator. All methods apart from the feature loss based CRN model manage to create high quality visual reconstructions of the \textit{monoE} images. 
Qualitative samples can be seen in Fig.~\ref{fig:comparison}. Areas around pulmonary embolisms are further highlighted.

\noindent\textbf{Impact on automatic PE diagnosis:}
\noindent The quantitative results on the classification results of a ResNet50 network trained on various input image domains are displayed in Table~\ref{tab:rnsa}. Despite the similar SSIM/PSNR results, the L1-loss-based model generates images, which slightly hamper the classification ability of a model. The other compared models worsen the performance, while our proposed method manages to generate visually fitting images as well as improves classification results over the baseline.
    \setlength{\abovecaptionskip}{1pt}
    \setlength{\belowcaptionskip}{50pt}
\begin{table}[t]
    \centering
    \begin{tabular}{|c||c||c|c||c|c|c||c|}
    \hline
    Domain & \small  Orig. & \small  L1& \small  SPL& \small  Pix2Pix& \small  CRN*& \small  P2PHD* & Ours\\\hline\hline
    AuROC & \small 0.8142& \small 0.8102& \small 0.8061& \small 0.8051& \small 0.8038& \small 0.8019 & \small \textbf{0.8420} \\\hline 

    \end{tabular}
    \caption{\textit{Pulmonary Embolism}-Classification results of a ResNet-50 trained on images from different image domains of various Image-Translation methods. Best result in \textbf{bold}.}
    \label{tab:rnsa}
\end{table}

\section{Discussion}\label{sec:Discussion}
We have investigated the potential use of the prediction of monoenergetic from polyenergetic images for the automatic identification of pathologies in CTPAs. We have displayed that most established image translation method either fail to correctly reconstruct that domain or are dismissing features necessary for classification. To offset these shortcomings of existing approaches we introduce an end-to-end learnable framework which combines the training of the classification and translation network. The reconstruction loss terms manage to let the network predict the visual properties, while the classification loss lets it enhance distinguishable features for the trained task. 
Results on the RSNA STR Pulmonary Embolism Detection dataset indicate that our approach provides a successful domain adaptation to monoenergetic imagery as it outperforms existing image-translation methods for paired data, while using the same or less parameters.

\if
Results on fastMRIdataset  indicate  that  our  approach  is  successful  in  traininga  reconstruction  algorithm  that  removes  aliasing  artifacts,achieving  comparable  performance  to  the  conventional  su-pervised learning approach that has access to fully-sampleddata, while outperforming traditional compressed sensing andparallel imaging.  

The same neural network architecture wasused for the self-supervised and supervised training.  

Whilethis  is  not  the  focus  of  this  study,  other  choices  of  neuralnetworks  are  possible  for  further  improvement.   

The  effectof different choices ofΛwas also studied, where a variable-density selection with sufficient cardinality was favored.Fig. 5:a) and b) shows the SSIM and NMSE values for the 380 slicestested.  Supervised DL-MRI approach performs slightly better thanthe proposed self-supervised DL-MRI approach, while both methodsreadily outperform CG-SENSE quantitatively.

In  many  scenarios,  acquisition  of  fully-sampled  data  ischallenging  due  to  physiological  and  physical  constraints.The lack of ground truth data hinders the utility of the super-vised learning approaches in these scenarios.

The proposedself-supervised approach relies only on available sub-sampledmeasurements.  While we have focused on MRI reconstruc-tion, the proposed approach naturally extends to other linearinverse  problems,  and  has  potential  applications  in  otherimaging modalities. \fi

\section{Conclusion}
\label{sec:Conclusion}
The proposed joint optimization strategy allows training of reconstruction of monoenergetic images without losing features necessary for the classification process. Our method, hereby, improves noticeably over straight forward classification, while outperforming existing methods.

\section{Acknowledgements}\label{sec:Acknowledgements}
The present contribution is supported by the Helmholtz Association  under  the  joint  research  school  “HIDSS4Health - Helmholtz  Information and Data Science School for Health”.

\bibliographystyle{IEEEbib}
\bibliography{strings,refs}

\begin{thebibliography}{10}

\bibitem{spectral}
Prabhakar Rajiah, Suhny Abbara, and Sandra~Simon Halliburton,
\newblock ``Spectral detector ct for cardiovascular applications,''
\newblock {\em Diagnostic and Interventional Radiology}, vol. 23, no. 3, pp.
  187, 2017.

\bibitem{myo_study}
Rachid Fahmi, Brendan~L Eck, Jacob Levi, Anas Fares, Amar Dhanantwari, Mani
  Vembar, Hiram~G Bezerra, and David~L Wilson,
\newblock ``Quantitative myocardial perfusion imaging in a porcine ischemia
  model using a prototype spectral detector ct system,''
\newblock {\em Physics in Medicine \& Biology}, vol. 61, no. 6, pp. 2407, 2016.

\bibitem{pe_study}
Jakob Weiss, Mike Notohamiprodjo, Malte Bongers, Christoph Schabel, Stefanie
  Mangold, Konstantin Nikolaou, Fabian Bamberg, and Ahmed~E Othman,
\newblock ``Effect of noise-optimized monoenergetic postprocessing on
  diagnostic accuracy for detecting incidental pulmonary embolism in
  portal-venous phase dual-energy computed tomography,''
\newblock {\em Investigative radiology}, vol. 52, no. 3, pp. 142--147, 2017.

\bibitem{pix2pix}
Phillip Isola, Jun-Yan Zhu, Tinghui Zhou, and Alexei~A Efros,
\newblock ``Image-to-image translation with conditional adversarial networks,''
\newblock in {\em Proceedings of the IEEE conference on computer vision and
  pattern recognition}, 2017, pp. 1125--1134.

\bibitem{wolterink2017deep}
Jelmer~M Wolterink, Anna~M Dinkla, Mark~HF Savenije, Peter~R Seevinck,
  Cornelis~AT van~den Berg, and Ivana I{\v{s}}gum,
\newblock ``Deep mr to ct synthesis using unpaired data,''
\newblock in {\em International workshop on simulation and synthesis in medical
  imaging}. Springer, 2017, pp. 14--23.

\bibitem{he2016deep}
Kaiming He, Xiangyu Zhang, Shaoqing Ren, and Jian Sun,
\newblock ``Deep residual learning for image recognition,''
\newblock in {\em Proceedings of the IEEE conference on computer vision and
  pattern recognition}, 2016, pp. 770--778.

\bibitem{rsna}
``{RSNA STR Pulmonary Embolism Detection},
  {\url{https://www.kaggle.com/c/rsna-str-pulmonary-embolism-detection/rules}},
  {Accessed: 2020-10-25},'' .

\bibitem{pix2pixhd}
Ting-Chun Wang, Ming-Yu Liu, Jun-Yan Zhu, Andrew Tao, Jan Kautz, and Bryan
  Catanzaro,
\newblock ``High-resolution image synthesis and semantic manipulation with
  conditional gans,''
\newblock in {\em Proceedings of the IEEE conference on computer vision and
  pattern recognition}, 2018, pp. 8798--8807.

\bibitem{crn}
Qifeng Chen and Vladlen Koltun,
\newblock ``Photographic image synthesis with cascaded refinement networks,''
\newblock in {\em Proceedings of the IEEE international conference on computer
  vision}, 2017, pp. 1511--1520.

\bibitem{spl}
M.~Saquib Sarfraz, Constantin Seibold, Haroon Khalid, and Rainer Stiefelhagen,
\newblock ``Content and colour distillation for learning image translations
  with the spatial profile loss,''
\newblock in {\em Proceedings of the 30th British Machine Vision Conference
  (BMVC)}, 2019.

\end{thebibliography}

\end{document}